\documentstyle[12pt]{article}
\begin{document}

\def\frak{\cal }
\def\Bbb{\bf }

\title{The information interpretation of quantum mechanics}
\author{Karl Svozil\\
 {\small Institut f\"ur Theoretische Physik,}
  {\small Technische Universit\"at Wien }     \\
  {\small Wiedner Hauptstra\ss e 8-10/136}
  {\small A-1040 Vienna, Austria   }            \\
  {\small e-mail: svozil@tuwien.ac.at}}
\date{ }
\maketitle

\begin{flushright}
{\scriptsize http://tph.tuwien.ac.at/$\widetilde{\;\;}\,$svozil/publ/interpret.tex,dvi,htm,ps,tex$\}$}
\end{flushright}

\begin{abstract}
In the information interpretation of quantum mechanics,
information is the most fundamental, basic entity.
Every quantized system is associated
with a definite discrete amount of information \cite{zeil-99}.
This information content remains constant at all times and
is permutated one-to-one throughout the system evolution.
What is interpreted as measurement is a particular type of information
transfer over a fictitious interface.
The concept of a many-to-one state reduction
is not a fundamental one but results from
the practical impossibility to reconstruct the original state after the measurement.
\end{abstract}

\subsection*{Information and the quantum}

In the following we take the position that information is the most fundamental concept in
understanding the quantum.
This approach has been recently investigated by
Zeilinger and Brukner \cite{zeil-99,zeil-bruk-99,zeil-bruk-99a} and
Summhammer \cite{sum-00a}.
It can be traced back to Schr\"odinger's
``catalogue of expectation values''  \cite{schrodinger},
many coffee-house conversations here in Vienna \cite{greenberger:pr2,summhammer-pr}
as well as to the writings of Brillouin \cite{brillouin1}, Gabor \cite{gabor}
and the late Landauer  \cite{landauer-89}, among others (e.g., \cite{maxwell-demon}).

In standard treatments, the quantum evolution is presented in a twofold manner.
(i) Inbetween measurements, there is a unitary and thus reversible one-to-one evolution.
(ii) The measurement itself is modeled irreversibly, many-to-one, which is associated
with the ``wave function collapse'' or ``reduction of the state vector.''

In what follows we suggest to extend the unitary evolution also to the measurement
process.
Thereby, we assume a uniform reversible one-to-one quantum evolution
which is not interrupted by measurements causing many-to-one reductions.

This amounts to suggesting that the concept of irreversible measurement (ii)
is no deep principle but originates in the practical inability
to reconstruct a particular quantum state.
Reconstruction may widely vary
with technological capabilities which often boil down to financial
commitments.

{\em Information, in particular information encoded into a  quantum system, is conserved, irrespective
of whether or not a ``measurement'' has taken place.}

\subsection*{Measurement apparatus as interface}

In what follows,
the measurement apparatus is  modeled by an {\em interface} between the observer and the
observed subsystem.
An interface is introduced as a theoretical entity forming the common boundary between two parts
of a system, as  well as a means of information exchange between those
parts.
By convention, one part of the of the system is called ``observer''
and the other part ``object.''
Both the observer and the object are embedded into one and the same system.
The interface should be thought of as merely an intermediate construction,
a ``scaffolding,'' capable of providing the necessary conceptual means.

One could quite justifyable ask
(and this question {\em has} indeed been asked by Professor Bryce deWitt),
``where exactly {\em is} the interface in a concrete experiment,
such as a spin state measurement in a Stern-Gerlach apparatus?''

We take the position here that the location of the interface very much depends
on the physical proposition which is tested and on the conventions assumed.
Let us consider, for example, a statement  like
{\em ``the electron spin in the $z$-direction is up.''}

In the case of a Stern-Gerlach device, one could
locate the interface at the apparatus itself.
Then, the information passing through the interface is identified with
the way the particle took.

\begin{figure}
\begin{center}
\unitlength 1.00mm
\linethickness{0.4pt}
\begin{picture}(115.67,30.00)
\put(5.00,10.00){\circle{10.00}}
\put(11.67,10.00){\line(1,0){9.67}}
\put(21.33,10.00){\line(1,0){5.00}}
\multiput(26.33,10.00)(1.23,0.07){2}{\line(1,0){1.23}}
\multiput(28.80,10.14)(1.22,0.12){2}{\line(1,0){1.22}}
\multiput(31.24,10.38)(0.81,0.11){3}{\line(1,0){0.81}}
\multiput(33.65,10.71)(0.60,0.11){4}{\line(1,0){0.60}}
\multiput(36.05,11.14)(0.47,0.10){5}{\line(1,0){0.47}}
\multiput(38.42,11.67)(0.39,0.10){6}{\line(1,0){0.39}}
\multiput(40.76,12.28)(0.39,0.12){6}{\line(1,0){0.39}}
\multiput(43.08,13.00)(0.33,0.12){7}{\line(1,0){0.33}}
\multiput(45.38,13.81)(0.28,0.11){8}{\line(1,0){0.28}}
\multiput(47.65,14.71)(0.25,0.11){9}{\line(1,0){0.25}}
\multiput(49.90,15.71)(0.22,0.11){10}{\line(1,0){0.22}}
\multiput(52.13,16.81)(0.22,0.12){10}{\line(1,0){0.22}}
\put(21.33,10.00){\line(1,0){2.92}}
\put(24.25,9.97){\line(1,0){2.83}}
\multiput(27.08,9.86)(1.37,-0.08){2}{\line(1,0){1.37}}
\multiput(29.83,9.70)(1.33,-0.12){2}{\line(1,0){1.33}}
\multiput(32.49,9.46)(0.86,-0.10){3}{\line(1,0){0.86}}
\multiput(35.06,9.16)(0.62,-0.09){4}{\line(1,0){0.62}}
\multiput(37.55,8.78)(0.60,-0.11){4}{\line(1,0){0.60}}
\multiput(39.95,8.35)(0.46,-0.10){5}{\line(1,0){0.46}}
\multiput(42.26,7.84)(0.45,-0.11){5}{\line(1,0){0.45}}
\multiput(44.49,7.27)(0.36,-0.11){6}{\line(1,0){0.36}}
\multiput(46.62,6.62)(0.34,-0.12){6}{\line(1,0){0.34}}
\multiput(48.68,5.92)(0.28,-0.11){7}{\line(1,0){0.28}}
\multiput(50.64,5.14)(0.23,-0.11){8}{\line(1,0){0.23}}
\multiput(52.52,4.29)(0.23,-0.12){11}{\line(1,0){0.23}}
\put(56.00,22.00){\line(2,-3){2.67}}
\put(58.67,3.67){\line(-3,-5){2.40}}
\multiput(57.33,20.00)(0.40,0.11){6}{\line(1,0){0.40}}
\put(59.75,20.63){\line(1,0){2.14}}
\multiput(61.89,20.74)(0.47,-0.10){4}{\line(1,0){0.47}}
\multiput(63.75,20.34)(0.20,-0.12){8}{\line(1,0){0.20}}
\multiput(65.33,19.41)(0.12,-0.13){11}{\line(0,-1){0.13}}
\multiput(66.64,17.96)(0.11,-0.22){9}{\line(0,-1){0.22}}
\multiput(67.67,16.00)(0.12,-0.16){11}{\line(0,-1){0.16}}
\multiput(68.94,14.22)(0.14,-0.12){12}{\line(1,0){0.14}}
\multiput(70.61,12.82)(0.23,-0.11){9}{\line(1,0){0.23}}
\multiput(72.71,11.80)(0.52,-0.11){7}{\line(1,0){0.52}}
\multiput(57.67,1.33)(0.64,-0.11){4}{\line(1,0){0.64}}
\put(60.24,0.87){\line(1,0){2.23}}
\multiput(62.46,0.92)(0.38,0.11){5}{\line(1,0){0.38}}
\multiput(64.35,1.48)(0.17,0.12){9}{\line(1,0){0.17}}
\multiput(65.90,2.55)(0.12,0.19){15}{\line(0,1){0.19}}
\multiput(67.67,5.33)(0.12,0.26){6}{\line(0,1){0.26}}
\multiput(68.38,6.89)(0.14,0.12){9}{\line(1,0){0.14}}
\multiput(69.66,7.96)(0.37,0.12){5}{\line(1,0){0.37}}
\multiput(71.52,8.54)(2.40,-0.10){2}{\line(1,0){2.40}}
\put(76.67,7.00){\framebox(17.67,5.67)[cc]{}}
\put(78.33,12.67){\framebox(14.33,11.00)[cc]{UP}}
\put(19.67,15.00){\framebox(20.33,4.67)[cc]{}}
\put(19.67,0.33){\framebox(20.33,4.67)[cc]{}}
\put(105.00,15.00){\circle*{1.33}}
\put(110.00,15.00){\circle*{1.33}}
\put(115.00,15.00){\circle*{1.33}}
\put(56.00,22.00){\line(-3,-2){4.33}}
\put(58.67,18.00){\line(-3,-2){4.33}}
\put(56.33,-0.33){\line(-2,1){4.50}}
\put(58.67,3.67){\line(-5,3){4.83}}
\put(1.50,13.50){\line(1,-1){7.00}}
\put(8.50,13.50){\line(-1,-1){7.00}}
\put(15.00,-5.00){\dashbox{2.33}(29.67,35.00)[cc]{}}
\put(49.00,-5.00){\dashbox{1.33}(13.00,35.00)[cc]{}}
\put(73.33,-5.00){\dashbox{1.33}(24.33,35.00)[cc]{}}
\end{picture}
\end{center}
\caption{Where exactly is the interface located?
Different dashed boxes indicate different possibilities to locate it.
\label{where-is-interface}}
\end{figure}
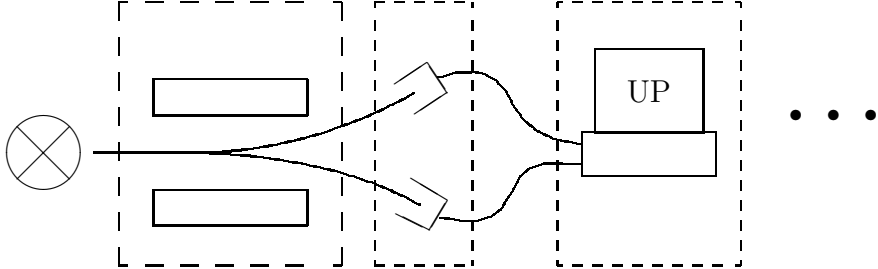

One could also locate the interface at two detectors at the end of the beam paths.
In this case, the ``meaningful'' informaton (with respect to the question asked)
penetrating through the interface corresponds
to which one of the two detectors (assumed lossless) clicks (cf. Fig. \ref{where-is-interface}).

The interface could also be situated at the computer interface card registering this
click, or at an
experimenter who presumably monitors the event
(cf. Wigner's friend \cite{wigner:mb}), or at the persons of the research group
to whom the experimenter reports, to their scientific peers, and so on.

It should be kept in mind that the proposition considered may be only a (minor) part of the information communicated
via the interface, and that in a uniform one-to-one environment,
{\em all} information would be needed for reconstruction.
As a consequence, it is certainly not sufficient to reconstruct the state of the electron.

The object may also not be prepared to accept the question asked.
In such a case, one may speculate that the interface acts as a ``translator''
between the observer and the object.
This may indeed be the reason of an intrinsic irreducible randomness of certain outcomes.
The quantum system as a whole behaves deterministically; i.e.,
according to the unitary evolution of the quantum state.

Since there is no material or real substrate which could be uniquely identified with
the interface, in principle it could be associated with or located at anything
which is affected
by the state of the object.
The only difference is the reconstructibility of the object's previous state (cf. below):
the ``more macroscopic'' (i.e., many-to-one) the interface is, the more difficult
it becomes to reconstruct the original state of the object.

\subsection*{Reconstruction and information flow density}
A direct consequence of the conservation of information
is the possibility to define continuity equations.
In analogy to magnetostatics or thermodynamics we may
represent the information flow by a vector which gives
the amount of information passing per unit area and per unit time through a surface element
at right angles to the flow. We call this
the {\em information flow density} ${\bf j}$.
The amount
of information flowing across a small area $\Delta A$ in a unit time is
$${\bf j}\cdot {\bf n}\; \Delta A,$$
where ${\bf n}$ is the unit vector normal to $\Delta A$.
The information flow density is related to the average flow velocity $v$ of information.
In particular, the information flow density associated with
an elementary object of velocity $v$ per unit time is given by
${\bf j}= \rho v$ bits per second, where $\rho $ stands for the
information density (measured in bits/$m^3$).
For $N$ elementary objects per unit volume carrying one bit each,
$${\bf j}= Nvi.$$
Here, $i$ denotes the elementary quantity of information measured in bit units.
The information flow $I$ is the total amount of information passing per unit time
through any surface $A$; i.e.,
$$I=\int_A {\bf j}\cdot {\bf n} \;dA.$$

We have assumed that the cut is on a closed surface ${\cal A}_c$ surrounding the object.
The conservation law of information requires the following continuity equation to
be valid:
$$\int_{{\cal A}_c}{\bf j}\cdot {\bf n}\; dA = -{d\over dt}({\rm Information\; inside})$$
or, by defining an information density $\rho$ and applying Gauss' law,
$$\nabla \cdot {\bf j}=   -{d\rho \over dt}.$$

To give a quantitative account of the present
ability to reconstruct the quantum wave function of single photons,
we analyze the ``quantum eraser'' paper by Herzog, Kwiat, Weinfurter and Zeilinger
\cite{hkwz}. The authors report an extension of their apparatus of $x= 0.13$~m,
which amounts to an information passing through a sphere of radius $x$ of
$$I_{\rm qe}= {4 \pi x^2 c\rho}=6\times 10^7 {\rm bits/second}.$$
Here, ${\bf j}=c\rho$ ($c$ stands for the velocity of light in vacuum) with
$\rho = 1 \textrm{bit/m}^3$ has been assumed.
At this rate the reconstruction of the photon wave function has been conceivable.

We propose to consider $I$ as a measure for wave function reconstruction.
In general, $I$ will be astronomically high because of the astronomical numbers of
elementary objects involved. Yet, the associated diffusion velocity $v$ may be considerably lower than $c$.

\subsubsection*{Effective many-to-one-ness}

In this final part of the communication we mention reasons why most measurements
appear to be irreversible.
In such cases, we claim, that either
(i) information flows off too fast, i.e., $I$ is too large, or
(ii) the interface is not total such that information ``leaks'' to regions
outside of the observer's control; or
(iii) the macroscopic level of description effectively maps
many different microscopic states onto a single macroscopic state.

The question of why on the macroscopic scale systems tend to behave irreversibly
while the
microphysical laws are reversible is not entirely new and has, in the context of statistical
mechanics, already been discussed
intensively by Boltzmann \cite{bricmont,maxwell-demon}.
In the case (iii), the interface effectively introduces ``classicality'' in the following sense.
The observer seeks an answer to a particular question or proposition.
This proposition is often only a part of the information the object is communicating.
In disregarding these other aspects, the experimenter induces
irreversibility into the particular description level.

In summary we have put forward here the suggestion
that irreversibility in quantum measurements is no primary concept.
It is postulated that, at least in principle, every quantum measurement could be ``undone.''


\end{document}